\documentclass[runningheads,a4paper]{llncs}

\usepackage{amsmath}
\usepackage{mathrsfs}
\usepackage{marvosym}
\usepackage{bbm}
\usepackage[utf8]{inputenc}
\usepackage{etex}
\usepackage{amssymb}
\usepackage{graphicx}
\usepackage{stmaryrd}
\usepackage[all]{xy}
\usepackage{color}
\usepackage{tikz-qtree}
\usepackage{bussproofs}
\usepackage[justification=centering,labelfont=bf]{caption}

\usepackage{url}
\urldef{\mailsa}\path{{grellois,mellies}@pps.univ-paris-diderot.fr}
\newcommand{\keywords}[1]{\par\addvspace\baselineskip
\noindent\keywordname\enspace\ignorespaces#1}

\newcommand{\iso}{\cong}

\newcommand{\data}{\texttt{data}}
\newcommand{\ift}{\texttt{if}}
\newcommand{\Nil}{\texttt{Nil}}

\newcommand{\G}{\mathcal{G}}

\newcommand{\valuetree}[1]{\langle #1 \rangle}
\newcommand{\valG}{\valuetree{\mathcal{G}}}

\newcommand{\runtree}[2]{\textbf{comp}(#1,#2)}
\newcommand{\tree}{\textit{witness}}
\newcommand{\leaves}[1]{\textbf{leaves}(#1)}

\newcommand{\interpA}[1]{ [\![ #1  ]\!]_{\mathcal{A}}}

\newcommand{\interpcol}[1]{ [\![\, #1 \, ]\!]}

\def\pole{\perp\!\!\!\perp}

\newcommand{\APT}{\mathcal{A}}

\newcommand{\superbang}{\lightning} %package stmaryrd

\newcommand{\colourbang}{\colorbang}
\newcommand{\colorbang}{\superbang\hspace{-.5765em}\superbang\hspace{-.5765em}\superbang}
\newcommand{\smallcolorbang}{\superbang\hspace{-.4765em}\superbang\hspace{-.4765em}\superbang}

\newcommand{\fixpoint}[1]{\textbf{Y}_{#1}}
\newcommand{\scottarrow}{\rightarrow}
\newcommand{\Scottcol}{\Scott_{\smallcolorbang}}
\newcommand{\FinScottcol}{\FinScott_{\smallcolorbang}}
\newcommand{\Scott}{\textbf{ScottL}}
\newcommand{\FinScott}{\textbf{FinScottL}}

\newcommand{\Pf}{\mathcal{P}_{fin}}

\begin{document}

\mainmatter 
\title{Finitary semantics of linear logic\\and higher-order model-checking}

%\titlerunning{Finitary semantics of linear logic\\and applications to higher-order model-checking}

\author{Charles Grellois \and Paul-Andr\'e Melli\`es}

\authorrunning{Charles Grellois \and Paul-Andr\'e Melli\`es}

\institute{Laboratoire PPS,\,Universit\'e Paris Diderot, Sorbonne Paris Cit\'e\\
\mailsa}

\maketitle

\begin{abstract}
In this paper, we explain how the connection between higher-order model-checking
and linear logic recently exhibited by the authors leads to a new and 
conceptually enlightening proof of the selection problem originally established
by Carayol and Serre using collapsible pushdown automata.
The main idea is to start from an infinitary and colored relational semantics
of the $\lambda\,Y$-calculus already formulated, and to replace it
by its finitary counterpart based on finite prime-algebraic lattices.
Given a higher-order recursion scheme $\mathcal{G}$, the finiteness of its interpretation
in the model enables us to associate to any MSO formula~$\varphi$ a new higher-order
recursion scheme~$\mathcal{G}_{\varphi}$ resolving the selection problem.

\end{abstract}

\abovedisplayskip = 4pt
\belowdisplayskip = 4pt
\abovedisplayshortskip = 4pt
\belowdisplayshortskip = 4pt

\section{Introduction}

Higher-order recursion schemes (HORS) provide an abstract model of computation
% which focuses on the flow of control induced by higher-order recursion.
which appears to be perfectly adapted for the task of model-checking functional programs.
Indeed, Knapik, Niwinski and Urzyczyn established in~\cite{knu} that for $n\geq 1$,
the trees generated by order-$n$ safe recursion schemes are exactly those 
that are generated by order-$n$ pushdown automata, and further, that they have
decidable MSO theories.
The MSO-decidability result for safe HORS was then extended a few years later
to all HORS by Ong \cite{ong}.
However, the MSO-decidability theorem established by the four authors
focuses on the decidability of a ``{local}'' model-checking problem:
\begin{quote}
Suppose given a HORS $\G$ which generates an infinite tree $\valG$.
Is it possible to decide for every MSO-formula $\varphi$ whether
the formula is valid at the root of the infinite tree~$\valG$.
\end{quote}
The MSO-decidability result means that the answer to this question is positive.
A more difficult ``global'' model-checking problem 
called the \emph{selection problem} in literature
%by Carayol and Serre \cite{...}
 is to understand whether:
\begin{quote}
Given a HORS $\G$ and a MSO-formula $\exists X \ \varphi[X]$
holding at the root of the infinite tree $\valG$, is it possible to compute 
a HORS $\G_{\varphi}$ generating a marked version $\valuetree{\G_{\varphi}}$
of the original tree $\valG$, and such that the set of its marked nodes 
is a witness $U$ satisfying the MSO-formula $\varphi[X]$.
\end{quote}
Quite strikingly, Carayol and Serre established in a recent paper~\cite{carayol-serre}
that the answer to this question is positive.
They also noticed that the selection problem follows from a purely automata-theoretic
property of HORS, which was established by Haddad in his PhD thesis~\cite{these-axel}:
\begin{quote}
Given a HORS $\G$ and an alternating parity tree automaton $\APT$ with the same ranked alphabet,
for every state~$q$ of the automaton~$\APT$ accepted by the tree~$\valuetree{\G}$,
it is possible to compute a HORS $\G_{q}$ generating an accepting run-tree $\valuetree{\G_q}$
of the automaton $\APT$ on the tree $\valG$ with initial state~$q$.
\end{quote}
Of course, the run-tree $\valuetree{\G_q}$ generated by the HORS $\G_{q}$
provides a witness of the fact that the state~$q$ is accepting.
But not only that: thanks to the equivalence between MSO-formulas
% on infinite trees
and alternating parity tree automata, the fact that the HORS $\G_q$
selects a \emph{specific} run-tree $\valuetree{\G_q}$ among all the run-trees with initial state~$q$
provides a solution to the ``selection problem''.
The idea is simply to extract from the run-tree $\valuetree{\G_q}$ a specific witness~$X$
for the MSO-formula $\exists X \ \varphi[X]$ satisfied by the tree $\valuetree{\G}$.

%
%Indeed, given a MSO-property~$\exists X \ \varphi[X]$ satisfied by the infinite tree~$\valG$,
%the set $X$ 

%
In this article, we will show how to establish the existence of such a ``higher-order recursive''
run-tree $\valuetree{\G_q}$ from purely denotational arguments,
%mixed with appropriate infinitary arguments.
based on a new and fundamental connection with linear logic developed by the authors
in a series of recent papers~\cite{fossacs,tensorial-logic-with-colours}.
In these papers, an infinitary and colored variant of the traditional semantics of linear logic
is constructed, see~\cite{fossacs} for details, and shown to compute in a compositional way
the set of accepting states of an alternating parity tree automaton, see~\cite{tensorial-logic-with-colours}
for details.
Despite the conceptual clarification this approach provides to higher-order model-checking,
this semantic account does not lead to any decidability result.
The reason is that the relational semantics of linear logic is a \emph{quantitative} semantics,
where finite types are interpreted as infinitary objects.
In order to establish decidability results, one thus needs to shift to  \emph{qualititative} semantics
where the interpretation of finite types remains finite.
This is precisely the purpose of the present paper: by shifting from the relational semantics
developed in~\cite{fossacs,tensorial-logic-with-colours} to the qualitative semantics
of linear logic provided by prime-algebraic lattices, we are able to establish
advanced decidability results like the theorem just mentioned by Carayol, Haddad and Serre.
This is the first time, to our knowledge, that such a strong and natural connection
between model-checking and the most contemporary tools of semantics
(linear logic, relational semantics) is exhibited.

\subsubsection*{\textbf{Plan of the paper.}} We start by recalling in \S\ref{section/hors}
the notion of higher-order recursion scheme and its correspondence with the $\lambda Y$-calculus. 
We then recall in \S\ref{section/homc} the notion of alternating parity tree automaton.
%as well as the selection problem.
In \S\ref{section/fincol}, we introduce a \emph{finitary} colored semantics of the $\lambda Y$-calculus, which
we use in \S\ref{section/type-denotations} to interpret $\lambda$-terms. We define a parameterized fixpoint in this model in \S\ref{section/rec},
obtaining colored semantics of the $\lambda Y$-calculus. In \S\ref{section/selec}, we use the finiteness of the model to prove the
decidability of the local model-checking and of the selection problem. We finally conclude in \S\ref{section/ccl}.

%\subsubsection*{\textbf{Related works.}}

%\begin{small}
%\subsubsection*{\textbf{Acknowledgments}:}
%The authors would like to thank Arnaud Carayol, Olivier Serre, Sylvain Salvati and Igor Walukiewicz
%for  discussions.
%\end{small}

\section{Higher-order recursion schemes and the $\lambda Y$-calculus}
\label{section/hors}

\noindent
\textbf{Higher-order recursion schemes.}  The set of simple types of the $\lambda$-calculus
is generated by the grammar
$\sigma,\,\tau \ ::=\ o \ \vert \ \sigma \rightarrow \tau$. We write $t\,::\,\sigma$
when a (possibly open) $\lambda$-term $t$ has 
simple type $\sigma$. Given a ranked alphabet $\Sigma$, a finite set of variables $\mathcal{V}$, 
a finite set of simply-typed
\emph{non-terminals} $\mathcal{N}$, and a distinguished non-terminal $S \in \mathcal{N}$, a higher-order recursion scheme (HORS) is the data, for every non-terminal $F \in \mathcal{N}$, of 
a closed simply-typed $\lambda$-term
\begin{equation}\label{eq/rf}
\mathcal{R}(F)\ \ =\ \ \lambda x_1.\ldots \lambda x_n.\,t
\end{equation}
of same type as the non-terminal~$F\in\mathcal{N}$, with constants in $\Sigma$,
where $x_i \in \mathcal{V}$ and $t\,::\,o$ is a $\lambda$-term of ground type without $\lambda$-abstractions.
Note that an element $a\in\Sigma$ of arity $n$ is represented as a constant
of type $o\rightarrow \cdots \rightarrow o \rightarrow o$ with same arity~$n$.
For each non-terminal~$F\in\mathcal{N}$, the data provided by $\mathcal{R}(F)$
is equivalently represented as a rewrite rule
$$
F\,t_1\,\ldots\,t_n \ \ \rightarrow_{\mathcal{G}}\ \ t[x_i \leftarrow t_i].
$$
Every higher-order recursion scheme $\mathcal{G}$ generates a potentially infinite 
$\Sigma$-labelled ranked tree noted~$\valG$ and called its \emph{value tree}.
This tree is simply obtained by applying an infinite number of times and in a fair way
the rewrite rules $\rightarrow_{\mathcal{G}}$ of the HORS $\mathcal{G}$
starting from the start symbol $S\in\mathcal{N}$.

\begin{example}
\label{example/valG}
Given $\Sigma\,=\,\{\,\texttt{if}\,:\,2,\,\texttt{data}\,:\,1,\,\texttt{Nil}\,:\,0\,\}$, consider the HORS $\mathcal{G}$
\begin{equation}\label{eq/hors}
\left\{
\begin{tabular}{rcl}
$\texttt{S}$ & $\quad = \quad $ & $\texttt{L Nil}$\\
$\texttt{L}$ & $\quad = \quad $ & $\lambda x.\,\texttt{if } x\ (\ \texttt{L }(\ \texttt{data } x\ ) \ )$\\
\end{tabular}
\right.
\end{equation}
which abstracts a simple program whose function $\texttt{Main}$ (abbreviated as $\texttt{S}$) calls a function \texttt{Listen} (denoted $\texttt{L}$),
starting from an empty list. 
Depending on a side condition unknown to the user or abstracted by the model-checker, 
\texttt{Listen} either returns a stack of data,
or receives a new element and pushes it on the current stack.
The value tree~$\valG$
%$\rightarrow_{\mathcal{G}}$
of this scheme, depicted in Figure~\ref{fig/value-tree},
provides an abstraction of the set of potential executions of the program.
Note that even though the program $\texttt{Main}$ is very simple, 
its execution tree $\valuetree{\mathcal{G}}$ is not regular, 
since it admits an infinite number of different subtrees.
This justifies from a practical point of view to study how traditional model-checking techniques
could be adapted to the HORS $\mathcal{G}$.

%
%The only finite representation of the value tree of a recursion scheme is therefore the recursion scheme itself. 
%This is the reason why semantics got so deeply involved in the proofs of decidability of the local model-checking problem.
\end{example}

\begin{figure}[t]
\small
\begin{minipage}{.5\textwidth}
$$
\begin{tikzpicture}
\tikzset{level distance=22pt}
\Tree [.$\ift$ $\Nil$ [.$\ift$ [.$\data$ $\Nil$ ] [.$\ift$ [.$\data$ [.$\data$ $\Nil$ ] ] $\vdots$ ] ] ]
\end{tikzpicture}
$$
\caption{An order-1 value tree.}
\label{fig/value-tree}
\end{minipage}
\begin{minipage}{.5\textwidth}
$$
\begin{tikzpicture}
\Tree [.$(\ift,\,q_0)$ [.$(\ift,\,q_0)$ [.$(\ift,\,q_0)$ $\vdots$ ] [.$(\ift,\,q_1)$ $\vdots$ ] ] [.$(\ift,\,q_1)$ [.$(\data,\,q_1)$ $\vdots$ ] [.$(\ift,\,q_0)$ $\vdots$ ] ] ]
\end{tikzpicture}
$$
\caption{An APT run-tree.}
\label{fig/run-tree}
\end{minipage}
\end{figure}

\noindent
\textbf{$\lambda$-calculus with recursion.}
It is well-known among the specialists of the $\lambda$-calculus
that higher-order recursion schemes can be nicely represented 
as simply-typed $\lambda$-terms in a $\lambda$-calculus
extended with a fixpoint operator~$Y$.
The resulting $\lambda Y$-calculus is thus defined by adding to the simply-typed $\lambda$-calculus,
a fixpoint operator $Y_{\sigma}$ of type $\sigma \rightarrow \sigma$
together with a rewriting rule
$$
Y_{\sigma} \ M \ \ \rightarrow_{\delta} \ \ M\ (\, Y_{\sigma} \ M \,)
$$
for every simple type $\sigma$.

\begin{proposition}
\label{prop/hors-lambdaY}
For every HORS $\mathcal{G}$ of ranked alphabet~$\Sigma$,
there exists a closed $\lambda Y$-term $t\,::\,o$ with constants in $\Sigma$,
such that the $\lambda Y$-term $t$ converges to the value-tree $\valG$
in the traditional sense of B\"ohm trees in the $\lambda Y$-calculus.
Conversely, there exists for every closed $\lambda Y$-term $t\,::\,o$ with constants in $\Sigma$
a HORS $\mathcal{G}$ of same ranked alphabet~$\Sigma$, such that 
the $\lambda Y$-term $t$ converges to~$\valG$.
\end{proposition}

\noindent
%The key idea is to translate a rule $F \,=\, \mathcal{R}(F)$ of type $\sigma$ as a $\lambda Y$-term $Y_{\sigma}\,(\lambda F.\,\mathcal{R}(F)\,)$,
%by duplicating the non-terminals in order to avoid mutual recursion.
An important benefit of this equivalence property is that the $\lambda Y$-calculus
is very well understood from the semantic point of view, and thus somewhat simpler
to study mathematically speaking than higher-order recursion schemes.
%
%Given a HORS $\G$, we denote $t(\G)$ the corresponding $\lambda Y$-term.
%An important consequence of this proposition is that it enables us to interpret HORS
%in denotational models of the $\lambda$-calculus equipped with a suitable recursion operator.

%\section{Higher-order model-checking and the selection problem}
\section{MSO and alternating parity tree automata}
\label{section/homc}
%
%\noindent
%\textbf{Monadic second-order logic.} Given a MSO formula $\varphi$ and a HORS $\G$,
%the \emph{local model-checking problem} consists in computing whether $\varphi$ holds at the root
%of $\valG$. It is known to be a decidable problem~\cite{ong}. When $\varphi\,=\,\exists X\ \psi[X]$,
%the \emph{MSO selection problem} is to find an algorithm
%% whose input is $(\G,\,\varphi)$ and 
%which outputs \texttt{false} if $\varphi$ is not
%satisfiable at the root of $\valG$, and else a new recursion scheme $\Gprim$ such that:
%\begin{itemize}
%\item the labels of $\valGprim$ are couples of a label of $\valG$ together with a boolean,
%\item the projection $\pi$ erasing booleans from labels is such that $\pi(\valGprim)\,=\,\valG$,
%\item the set $X$ of nodes marked with the boolean \texttt{true} in $\valGprim$ is such that $\psi[X]$ holds at the root of $\valG$.
%\end{itemize}
%In other terms, the algorithm computes a recursion scheme producing a lifting of $\valG$ with booleans marking a set satisyfing $\psi$, if any.
%Note that the selection problem is stronger than the \emph{global model-checking problem} which, given a MSO formula $\chi$,
%consists in finding a recursion scheme whose value tree is marked with booleans indicating precisely the nodes where the formula holds.
%It suffices indeed to consider the selection problem for the MSO formula $\psi[X]\,=\,x \in X \Leftrightarrow \chi(x)$. In general, selection is
%strictly harder than global model-checking~\cite{hdr-olivier}.\\
%
%
%\noindent
%\textbf{Alternating parity automata.}

As explained in the introduction, there is a beautiful correspondence
between the formulas of monadic second-order logic (MSO) 
and alternating parity tree automata, which we briefly recall here
for the sake of completeness.

\begin{proposition}
\label{prop/equiv-MSO-APT}
For every ranked alphabet~$\Sigma$, one has the following equivalence:
%, $\varphi$ a MSO formula over $\Sigma$-labelled trees, and $\APT$ an
%alternating parity tree automaton (APT) over $\Sigma$.
\begin{itemize}
\item Every MSO formula~$\varphi$ over $\Sigma$-labelled trees can be translated 
to an APT $\mathcal{A}_{\varphi}$ of same ranked alphabet $\Sigma$, 
such that $\varphi$ holds at the root of a $\Sigma$-labelled tree~$T$
iff $\mathcal{A}_{\varphi}$ has an accepting run-tree over $T$
from its initial state~$q_0$.
\item Conversely, every APT $\mathcal{A}$ of ranked alphabet~$\Sigma$ can be translated 
to a MSO formula $\varphi_{\mathcal{A}}$ of same ranked alphabet, such that for every $\Sigma$-labelled
tree $T$, $\mathcal{A}$ has an accepting run-tree over $T$ from its initial state~$q_0$
if and only if the MSO-formula $\varphi_{\mathcal{A}}$ holds at the root of $T$.
\end{itemize}
\end{proposition}
Recall that alternating parity tree automata (APT) are non-deterministic top-down tree automata with the additional ability to \emph{duplicate} or to \emph{erase} subtrees.
Typical transitions are thus of the form
\begin{equation}\label{eq/APT}
\delta(q_0,\texttt{if}) = (2,q_0)\wedge (2,q_1) \quad
\quad\quad
\quad \delta(q_1,\texttt{if})=(1,q_1)\wedge (2,q_0)
\end{equation}
When a node labelled with $\texttt{if}$ is visited in state $q_0$, its left subtree 
is ``dropped'' or ``erased'' while the right one is ``explored twice'' or ``duplicated'',
with $q_0$ as initial state in one copy, and $q_1$ as initial state in the other copy.
The second transition does not use alternation, and
would be usually written as $(q_1,\,\texttt{if},\,q_1,\,q_0) \in \Delta$
in a nondeterministic tree automaton.
Run-trees of an alternating parity tree automaton are unranked,
and their shape may differ a lot from the original tree. 
The effect of the transitions (\ref{eq/APT}) over the tree of Figure \ref{fig/value-tree}
is depicted in Figure \ref{fig/run-tree}. In general, a transition is of the shape
\begin{equation}\label{equation/apt-transition}
\delta(q,\,a) \ \ =\ \ \bigvee_{i \in I} \ \bigwedge_{j \in J_i} (d_{i,j}, \, q_{i,j}) \ \ =\ \ \bigvee_{i \in I}\ \varphi_i
\end{equation}
where the union stands for non-determinism, and the conjunction for alternation: after $i$ is chosen, for every $j \in J_i$,
the automaton runs with state $q_{i,j}$ over a copy of its subtree in direction $d_{i,j}$. 
For every $i$, we say that $\varphi_i$ is a \emph{conjunctive clause} of the formula $\delta(q,a)$.

Seen from an automata-theoretic point of view, monadic second-order (MSO) logic
is equivalent to the modal $\mu$-calculus.
As such it enables one to express safety properties 
(typically, that a given state ``error'' is never encountered) as well as
liveness properties (typically, that a given state ``happy'' is visited infinitely often).
The safety properties are inductive: it is enough to check that no finite approximation
of a computation enters an error state, while the liveness properties are coinductive, 
since they specify infinitary behaviors.
Moreover, MSO logic and the modal $\mu$-calculus are sufficiently expressive
to alternate these inductive and coinductive specifications.
This alternation is handled by extending APT
with a \emph{parity condition} over their run-trees.
Alternating parity automata are thus equipped with 
a \emph{coloring function} $\Omega\,:\,Q \rightarrow \mathbb{N}$,
which associates a color to each state $q$ of the automaton.
This coloring of the states $q\in Q$ of the automaton induces 
a coloring of the nodes of its run-trees, in the expected way.
Following the principles of parity games, an infinite branch of such a run-tree is declared winning
when the greatest color occurring infinitely often in it is even.
A run-tree of the automaton is then accepted precisely when all its infinite branches are winning.
In the sequel, we find convenient to consider the set $Col\ =\ \Omega(Q) \uplus \left\{\epsilon\right\}$
of colors appearing in the alternating parity automaton $\mathcal{A}$ under study.
The extra color $\epsilon$ is added as a neutral color, in order to reflect 
the comonadic nature of colors, as we will explain in the later~\S\ref{section/fincol}.
The following definition will also be useful in the sequel, in order to connect
the alternating parity automaton $\mathcal{A}$ and the finitary semantics of linear logic:

\begin{definition}
Given a state $q\in Q$ and a $n$-ary constructor $a \in \Sigma$,
we say that a $n$-tuple $\alpha \in \left(\Pf(Col \times Q)\right)^n$
satisfies the formula $\delta(q,a)$ when $\alpha$ is of the form
$$
\alpha \ \ =\ \ \left( \,\, \left\{\,(c_{1i_1},\,q_{1i_1})\ \vert \ i_1 \in I_1\right\} \,\, ,\,\,\ldots
\,\, , \,\,\,\left\{\,(c_{ni_n},\,q_{ni_n})\ \vert \ i_n \in I_n\right\}\,\,\right)
$$
and there exists a $n$-tuple of subsets $J_1 \subseteq I_1$, \ldots, $J_n \subseteq I_n$ 
such that
\begin{equation}\label{equation/refinement}
\bigwedge_{k=1}^n\ \bigwedge_{j_k \in J_k} \ \left(k,\,q_{kj_k}\right)
\end{equation}
defines a conjunctive clause of the formula $\delta(q,a)$,
and such that moreover
$$
\forall k \in \{1,\ldots,n\}\ \ \forall j \in J_k\ \ c_{kj}\,=\,\Omega(q_{kj}).
$$
\end{definition}

\noindent
In other words, $\alpha$ is a $n$-tuple of sets $\left\{(c_{1i_k},\,q_{1i_k})\ \vert \ i_k \in I_k\right\}$
of states annotated with colors, 
each of them corresponding to one of the $n$ subtrees below the symbol $a$.
Moreover, each such set should contain
a subset $\left\{(\Omega(q_{1i_k}),\,q_{1i_k})\ \vert \ i_k \in J_k\right\}$ of appropriately
colored states, such that (\ref{equation/refinement}) defines a conjunctive clause of the formula $\delta(q,a)$.
The general idea is that the $n$-tuple is allowed to contain more colored states
than what is stricly required for the transition $\delta(q,a)$ to be performed
by the alternating parity automaton~$\mathcal{A}$.
%each direction $k\in\{1,\dots,n\}$ enough colored states
%give a set satisfying $\delta(q,\,a)$, and are properly colored by~$\Omega$.
%But $\alpha$ contains ``more states than required by $\delta$'' to run successfully from $a$ in state $q$;
%and there is no coloring restriction over the extra states.
This definition will be crucial in the construction of the finitary semantics which,
we will see, is based on downward-closed sets and subtyping.\\

\section{The Scott semantics of linear logic}
%A qualitative colored semantics of linear logic}
\label{section/fincol}
Here, we adapt the infinitary and colored relational semantics of linear logic
formulated in~\cite{fossacs,tensorial-logic-with-colours} to the finitary Scott semantics,
where formulas of linear logic are interpreted as partial orders.
%prime-algebraic lattices.
%
%
The semantics of linear logic is \emph{qualitative} in the technical sense that
its exponential modality~$!$ is interpreted using the finite \emph{powerset} construction,
which transports finite sets into finite sets, in contrast to the finite multiset construction
used in the traditional and \emph{quantitative} relational semantics.
The terminology of Scott semantics comes from the fact that 
in the derived semantics of the simply-typed $\lambda$-calculus,
every type is interpreted as a prime algebraic complete lattice,
and every simply-typed $\lambda$-term as a Scott-continuous function.
%
%However, we find convenient to develop here the formulation of linear logic
% due to Winskel~\cite{winskel}, and 
%obtained from the former by Stone duality, see for example~\cite{terui}.
So, let $\Scott$ denote the category with preorders $\mathbf{A}=(\,A,\,\leq_A\,)$ as objects
and downward-closed binary relations $R \subseteq A \times B$ as morphisms 
$(\,A,\,\leq_A\,) \rightarrow (\,B,\,\leq_B\,)$.
Here, by a downward-closed relation, we mean a binary relation~$R$ 
such that for all $a,a'\in A$ and $b,b'\in B$, one has :
$$
\left(a,\,b\right) \in R \hspace{.5em} \mbox{ and } \hspace{.5em} a\leq_A a' 
\hspace{.5em} \mbox{ and } \hspace{.5em} b' \leq_B b 
\hspace{2em} 
\Rightarrow \hspace{2em} (\,a',\,b'\,) \in R.
$$
The binary relation $R$ is thus downward closed in the partial order ${(A,\leq_A)}^{op}\times{(B,\leq_B)}$
interpreting the formula $(A,\leq_A)\multimap (B,\leq_B)$ in the Scott semantics.
The intuition guiding this property is that if a binary relation~$R$ interpreting a proof of linear logic
can produce an output $b$ from an input $a$, then the same binary relation
can also produce a less informative output $b'$ from a more informative input $a'$.
It is well-known in the literature on linear logic 
that this ``saturation property'' is essential in order to obtain 
a relational semantics of linear logic
with a qualitative (that is, based on finite sets instead of finite multisets)
interpretation of the exponential modality.
This remark is generally attributed to Ehrhard, see \cite{models-of-linear-logic} for details.
The composition in $\Scott$ is relational, since relational composition
preserves the property of being downward-closed.
The identity morphism over $(\,A,\,\leq_A\,)$ is
$$
id_A \ \ = \ \ \left\{\,(\,a',\,a\,)\ \vert \ a \leq_A a' \,\right\}
$$
$\Scott$ is a compact closed category with products, with
$$
\begin{tabular}{rclp{1.3cm}rcl}
$(\,A,\,\leq_A\,)\ \otimes\ (\,B,\,\leq_B\,)$ &  $\ \ =\ \ $ & $(\,A \times B,\,\leq_A \times \leq_B\,)$ &
 \quad &
$1$ &   $\ \ =\ \ $  & $(\,\{\star\},\,=\,)$\\
$(\,A,\,\leq_A\,)\ \&\ (\,B,\,\leq_B\,)$ &  $\ \ =\ \ $ & $(\,A \uplus B,\,\leq_A \uplus \leq_B\,)$
 &
\quad &
$\top$  &  $\ \ =\ \ $  & $(\,\emptyset,\,\emptyset\,)$ \\
$(\,A,\,\leq_A\,)^{\bot}$ & $\ \ =\ \ $ & $(\,A,\,\geq_A\,)$ & & & & \\
\end{tabular}
$$
The exponential modality 
$$! \quad : \quad A \quad \mapsto \quad {!A} \quad : \quad \Scott \quad \longrightarrow \quad \Scott$$
is then defined by associating to the ordered set~$(A,\leq_A)$ the set $\Pf(A)$ of finite subsets of~$A$,
where two finite subsets~$u$ and $v$ are ordered in the following way:
$$
u\leq_{!A} v \quad \quad \iff \quad \quad \forall a\in u, \, \exists b\in v, \quad u \leq_{A} v.
$$
%
%$R \subseteq A \times B$ to $\Pf (R) \subseteq \Pf(A) \times \Pf(B)$ defined as
%$$
%X \ \ \Pf R \ \ Y \quad \iff \quad \forall \alpha \in X\ \ \exists \beta \in Y \ \ \alpha \ R\ \beta
%$$
%
Recall that the endofunctor $!$ is transports every morphism $R\,:\,A \rightarrow B$
of the category~$\Scott$ to the following morphism:
%$$
%!\,A\ \ =\ \ (\,\Pf (A),\,\Pf \leq_A\,)
%$$
%\vspace{-0.65cm}
$$
!\,R\ \ =\ \ \left\{\,\left(u,\,v\right) \ \in\ !\,A\, \times\, !\, B \ \ \vert\ \ \forall\, b \in v \ \, \exists\, a \in u
\ \, \left(a,\,b\right) \in R \,\right\} \ \ :\ \ !\,A \rightarrow\ !\,B
$$ 
The endofunctor $!$ is in fact a comonad and defines a Seely category, 
and thus a model of full propositional linear logic,
based on the category~$\Scott$, see for instance~\cite{terui}.
%can be endowed with a monoidal comonad structure together with Seely isomorphisms, 
%, so that $\Scott$ is a model of linear logic.

\medbreak

\noindent
\textbf{The coloring comonad.} 
As we have shown in~\cite{fossacs,tensorial-logic-with-colours},
the treatment of colors by alternating parity automata follows essentially
the same comonadic principles as the treatment of copies in linear logic.
This connection between higher-order model checking and linear logic
leads to a coloring comonad $\Box$ on the relational semantics of linear logic,
which we adapt here to the qualitative Scott semantics.
To that purpose, we fix a finite set of colors $Col$ containing a neutral element~$\epsilon$,
and consider the coloring function $Q\to Col$ which associates a color to every state
of a parity tree automaton~$\mathcal{A}$, see the previous discussion in~\S\ref{section/homc}.
The modality~$\Box$ is then defined in the following way for an ordered set~$(A,\leq_A)$
and a morphism ${R:(A,\leq_A)\to(B,\leq_B)}$:
$$
\begin{tabular}{rcl}
$\square \ (\,A,\,\leq_A\,) $ & $ \ \ = \ \ $ & $ (\,A,\,\leq_A\,\,)\, \& \, \cdots \, \& \, (\,A,\,\leq_A\,)$ \\
& $\ \ \iso\ \ $ & $  \left( \left\{(i,\,a)\ \vert\ i \in Col,\,a \in A \right\},\, \leq_{\square\,A} \, \right) $\\
$(i,\,a)\ \ \square\, R\ \ (j,\,b) $ & $ \mbox{ iff }$ & $ i\,=\,j \mbox{ and } a\, R\, b$\\
\end{tabular}
$$
where $(i,\,a) \leq_{\square\,A} (j,\,a')$ iff $i=j$ and $a \leq_A a'$.
The comonadic structure of $\Box$ is provided by the following structural morphisms
$$
\begin{tabular}{rclcl}
$\mathbf{dig}_A$ & $\ = \ $ & $\left\{ ( ( \operatorname{max}(c_1, c_2), a) , ( c_1, ( c_2, a' )  ) )\ \vert \ a' \leq_A a  \right\}$ & $\,:\, $ & $\square A \scottarrow \square \square A$\\
$\mathbf{der}_A$ & $\ = \ $ & $\left\{ ( ( \epsilon, a ), a' ) \ \vert \ a' \leq_A a   \right\}$ & $\, :\, $ & $\square  A \scottarrow A$\\
$m_{A,B}$ & $\ = \ $ & $\left\{ ( ( ( i, a ), ( i, b ) ), ( (i, (a', b' ) ) ) ) \ \vert \ a' \leq_A a,\, b' \leq_B b   \right\}$ & $\, :\, $ & $\square  A \otimes \square  B \scottarrow \square ( A \otimes B )$\\
$m_1$ & $ \ =\ $ &$\left\{\, (\star,\,(c,\,\star))\ \vert \ c \in Col\, \right\}$ & $\, :\, $ & $1 \scottarrow \square\, 1$\\
%$\laxzerocol$ & $\ =\ $ & $\left\{(- \right\}$ & $\,:\,$ & $1 \scottarrow \square 1$\\
\end{tabular}
$$
As we did in the case of the relational semantics~\cite{fossacs,tensorial-logic-with-colours},
we define a distributive law $\lambda\,:\,{!}\circ{\square} \Rightarrow {\square}\circ{!}$
between the comonads $!$ and $\square$ defined as the natural transformation:
$$
\lambda_A\ =\ \left\{\left(\left\{\left(c_j,\,a'_j\right)\right\},\left(c,\left\{ a_i \right\} \right) \right) \ \vert\ \forall\, i\ \exists\,j\ \ c=c_j \mbox{ and }a_i \leq_A a'_j  \,\right\}\ :\ !\,\square\,A \rightarrow \square\,!\,A
$$
The existence of such a distributive law~$\lambda$ enables us to equip
the composite functor $\colorbang \ =\ {!}\circ{\square}$ with a comonadic structure.
It appears moreover that this colored exponential functor $\colorbang$ 
satisfies the axioms of a Seely category, and thus defines a model of full propositional linear logic.
We denote by $\Scottcol$ its Kleisli category.
%, which is cartesian closed and is thus a model of the $\lambda$-calculus.

\section{A finitary interpretation of the simply-typed $\lambda$-calculus}
\label{section/type-denotations}

In order to simplify the discussion, we suppose given an alternating parity tree~$\APT$ 
over a signature $\Sigma$, with set of states $Q$ and with transition function $\delta$.
As a Kleisli category associated to a model of linear logic,
the category $\Scottcol$ is cartesian closed and thus a model 
of the simply-typed $\lambda$-calculus.
The simple types are interpreted inductively as
$$
\interpcol{\sigma \rightarrow \tau}\ \ =\ \ \colorbang\,\interpcol{\sigma} \multimap \interpcol{\tau}
\quad \mbox{ and } \quad
\interpcol{o}\ \ =\ \ \pole \ \ =\ \ (\,Q,\,=\,)
$$
%
%Notice that the linear arrow $\multimap$ reverses the order on $\colourbang\,\interpcol{\sigma}$. The ordering relation over simple types can be presented as a subtyping  system, see Figure~\ref{fig/order}.
%as done in particular by Terui \cite{terui}.\\
%
The interpretation of the simply-typed $\lambda$-terms is standard,
except for the interpretation of the elements of the ranked alphabet~$\Sigma$,
seen as here constants of the simply-typed $\lambda$-calculus,
which are interpreted as follows:
$$
\interpA{\,a\,} \ \ = \ \ \{\,(\alpha,q)\ \vert \ q \in Q \mbox{ and  } \alpha \mbox{ satisfies the formula } \delta(q,a) \,\}
$$
As explained in~\cite{tensorial-logic-with-colours} in the case of the quantitative relational semantics
of linear logic, this interpretation of the elements of $\Sigma$ corresponds to a Church encoding 
of the alternating parity automaton~$\APT$, encoded in the present case 
in the qualitative Scott semantics of linear logic.

\begin{example}
Recall the two transitions (\ref{eq/APT}) introduced as running example in \S\ref{section/homc}:
$$\delta(q_0,\texttt{if}) = (2,q_0)\wedge (2,q_1) \quad
\quad\quad
\quad \delta(q_1,\texttt{if})=(1,q_1)\wedge (2,q_0)
$$
Setting $c_i\,=\,\Omega(q_i)$, these transitions imply that
$$
\left(u_1,\,u_2,\,q_0\right) \in \interpA{if} \quad \mbox{ and } \quad 
\left(v_1,\,v_2,\,q_1\right) \in \interpA{if}
$$
for every finite sets $u_1,\,u_2,\,v_1,\,v_2 \in \colorbang \!\!\pole \ =\ \Pf \left(Col \times Q\right)$
satisfying moreover that $\left\{(c_0,\,q_0),\,(c_1,\,q_1)\right\} \subseteq u_2$,
that $(c_1,\,q_1) \in v_1$ and that $(c_0,\,q_0) \in v_2$.
\end{example}

%
%
%
%Consider a context of simple types for the free variables of a $\lambda$-term $t$,
%proving that it admits the simple type $\tau$
%$$
%\Gamma \ \ =\ \ x_1\,::\,\sigma_1,\,\cdots \,,x_n\,::\,\sigma_n \ \ \vdash \ \ t\,::\,\tau
%$$
%%
%We define the denotation of $t\,::\,\tau$ in the context $\Gamma$ as
%$$
%\interpcol{\Gamma\ \vdash\ t\,::\,\tau}\ \ \subseteq \ \ \left(\,\colourbang\,\interpcol{\sigma_1} \otimes \cdots \otimes \colourbang \interpcol{\sigma_n}\,\right) \multimap \interpcol{\tau}
%$$
%by structural induction over terms:
%$$
%\begin{tabular}{rcl}
%$\interpcol{\emptyset \ \vdash \ a\,::\,o \rightarrow \cdots \rightarrow o}$ & $\ \ = \ \ $&$ \{\,S \ \vert \ S \mbox{ refines } \delta \mbox{ from } a\,\} $\\
%$\interpcol{x\,::\,\sigma \ \vdash \ x\,::\,\sigma}$ &$ \ \ = \ \ $& $\left\{\,X \subseteq \sigma  \right\}$\\
%$\interpcol{\Gamma\ \vdash \ \lambda x.\,M\,::\,\sigma \rightarrow \tau}$ &$ \ \ = \ \ $& \\
%$\interpcol{\Gamma\ \vdash \ M\ N\,::\,\sigma}$ &$ \ \ = \ \ $& \\
%\end{tabular}
%$$
%For a closed term $t$ of simple type $\sigma$, we set $\interpcol{t}\,=\,\interpcol{\emptyset\ \vdash\ t\,::\,\sigma}$.
%% TODO : ici, clôture des dénotations
%
%
%

%\medbreak

%\noindent
%\textbf{Denotations and intersection types.} 
\noindent
Using these interpretations in $\Scott$ of the elements of the ranked alphabet $\Sigma$, 
we construct the interpretation
$$
\interpA{\Gamma\ \vdash\ t\,::\,\tau}\ \ \subseteq \ \ \left(\,\colourbang\interpcol{\sigma_1} \otimes \cdots \otimes \colourbang \interpcol{\sigma_n}\,\right) \multimap \interpcol{\tau}
$$
of any $\lambda$-term $t$ of type~$\tau$ in a context of typed variables $\Gamma$,
with constants in the ranked alphabet~$\Sigma$.
An alternative way to describe this interpretation is to express it 
as an intersection type system with subtyping, in the style of 
Coppo, Dezani, Honsell and Longo~\cite{coppo-et-al} and 
more recently Terui~\cite{terui} in the framework of linear logic.
In this formulation, sequents are of the following form
$$
\Gamma\ \ =\ \ x_1\,:\,u_1\,::\,\sigma_1,\,\ldots,\,x_n\,:\,u_n\,::\,\sigma_n \ \ \vdash \ \ t\,:\,\alpha\,::\,\tau
$$
where $u_i \in \colourbang \interpcol{\sigma_i}$ and $\alpha \in \interpcol{\tau}$. 
The typing rules are presented in Figure~\ref{fig/typesys},
with the subtyping relation $\leq_A$ defined inductively in Figure~\ref{fig/order}.
Note that the coloring  $\square_c\ \Gamma$ of a context is defined inductively as
$$
\begin{tabular}{rcl}
$\square_c\, \left(x\,:\,u\,::\,\sigma,\,\Gamma \right)$
& $\ \  = \ \ $
& $ x\,:\,\square_c\ u\,::\,\sigma,\ \square_c\,\Gamma $\\
$\square_c\,\left\{\,(c_i,\,\alpha_i)\,\right\}$
&$\ \  =\ \ $
& $ \left\{\,(\operatorname{max}(c,c_i),\,\alpha_i)\,\right\} $\\
\end{tabular}
$$
\begin{figure}[t]
\centering
\begin{tabular}{ccccc}
$$
\AxiomC{}
\UnaryInfC{$q \leq_{\pole} q$}
\DisplayProof
$$

&

\hspace{0.8cm}

&

$$
\AxiomC{$\forall \, (c,\,\alpha) \in u\ \ \exists\,(c,\, \beta) \in v \ \ \alpha \,\leq_A\, \beta$}
\UnaryInfC{$u\,\leq_{\smallcolorbang A} \, v$}
\DisplayProof
$$
&
\hspace{0.8cm}
&
$$
\AxiomC{$v\, \leq_{\smallcolorbang A} \, u $}
\AxiomC{$\alpha \,\leq_B\,\beta$}
\BinaryInfC{$u \rightarrow \alpha\, \leq_{\smallcolorbang A \multimap B} \, v \rightarrow \beta $}
\DisplayProof
$$

\end{tabular}

\caption{Inference rules for the preorders associated with simple types.}
\label{fig/order}
\end{figure}

\begin{figure}[t]

 \makebox[\textwidth][c]{
\begin{tabular}{ccccc}
$$
\AxiomC{$\exists\ (\epsilon,\,\alpha') \in  u\ \ \ \alpha\ \leq_{\interpcol{\sigma}}\, \alpha'$}
\LeftLabel{Ax \ \ }
%\RightLabel{}
\UnaryInfC{$x\,:\,u\,::\,\sigma \ \vdash\ x\,:\,\alpha\,::\,\sigma$}
\DisplayProof
$$
&
\hspace{0.5cm}
&

$$
%\AxiomC{$\alpha \in \interpcol{\emptyset \ \vdash \ a\,::\,\sigma}$}
\AxiomC{$q \in Q$ and $\alpha$ satisfies $\delta(q,a)$}
\LeftLabel{$\delta$ \ \ }
%\RightLabel{}
\UnaryInfC{$\emptyset \ \vdash\ a\,:\,\alpha \rightarrow q\,::\,\sigma$}
\DisplayProof
$$

&
\hspace{0.5cm}
&
$$
\AxiomC{$\Gamma,\, x\,:\,u\,::\,\sigma \ \vdash \ M\,:\,\alpha\,::\,\tau$}
\RightLabel{\ \ $\lambda$}
%\RightLabel{}
\UnaryInfC{$\Gamma \ \vdash \ \lambda x.\,M\,:\,u \rightarrow \alpha\,::\,\sigma \rightarrow \tau$}
\DisplayProof
$$\\
\end{tabular}
}

$$
\AxiomC{$\Gamma_0 \ \vdash \ M\,:\,\{\,(c_1,\, \beta_1),\,\ldots,\,(c_n,\, \beta_n)\,\} \rightarrow \alpha\,::\,\sigma \rightarrow \tau$}
\AxiomC{$\Gamma_i \ \vdash \ N\,:\,\beta_i\,::\,\sigma \quad \mbox{(for all } i\mbox{)}$}
\LeftLabel{App \quad}
\BinaryInfC{$\Gamma_0 \cup \square_{c_1}\, \Gamma_1 \cup \cdots \cup \square_{c_n} \, \Gamma_n \ \vdash \ M\, N \,:\,\alpha\,::\,\tau$}
\DisplayProof
$$

\caption{Type-theoretic computation of denotations in $\Scottcol$}
\label{fig/typesys}
\end{figure}

\begin{proposition}
\label{prop/type}
The sequent 
$$
\Gamma\ =\ x_1\,:\,u_1\,::\,\sigma_1,\,\ldots,\,x_n\,:\,u_n\,::\,\sigma_n \ \vdash \ t\,:\,\alpha\,::\,\tau
$$
is provable in this intersection type system if and only if
$$
(u_1,\,\ldots,\,u_n,\,\alpha) \ \in \ \interpA{\Gamma\ \vdash\ t\,::\,\tau}  \ \subseteq \ \left(\,\colourbang\,\interpcol{\sigma_1} \otimes \cdots \otimes \colourbang \interpcol{\sigma_n}\,\right) \multimap \interpcol{\tau}
$$
\end{proposition}

\noindent
%For a closed term $t$ of simple type $\sigma$, we set $\interpA{t}\,=\,\interpA{\emptyset\ \vdash\ t\,::\,\sigma}$.
%A consequence of Proposition \ref{prop/type} is that the typing derivations describe the computations of the elements of the denotation of a term.

\section{The recursion operator~$Y$}
\label{section/rec}
At this stage, we are ready to shift from the colored semantics of the simply-typed $\lambda$-calculus
formulated in \S\ref{section/type-denotations}
to a colored semantics of the simply-typed $\lambda Y$-calculus.
To that purpose, we construct a Conway operator~$\mathbf{Y}$ in the category $\FinScott$
defined as the full subcategory of $\Scott$ consisting of the \emph{finite} ordered sets.
Note that $\FinScott$ defines a Seely category, and thus a model of full propositional linear logic.
The Conway operator $\mathbf{Y}$ is defined a family of operations $\fixpoint{X,A}$ transporting 
a binary downward-closed relation
$$
R \ \ :\ \   \colorbang X \otimes \colorbang A \multimap A
$$
into a binary downward-closed relation
$$
\fixpoint{X,A}(R) \ \ :\ \ \colorbang X \multimap A
$$
and satisfying a series of conditions originally
stated by Bloom and Esik~\cite{bloom-esik}
in cartesian closed categories, and adapted in~\cite{fossacs} 
to the particular framework of Seely categories.
Note that, such a Conway operator on $\FinScott$ defines a Conway operator 
in the sense of~\cite{bloom-esik} in the cartesian-closed category $\FinScottcol$.
Just as in the case of the relational semantics, see \cite{fossacs} for details,
the important point here is that the colors added to the original Scott semantics
will enable us to alternate least and greatest fixpoints (and thus inductive
and coinductive reasoning) in the definition of the fixpoint operator~$\mathbf{Y}$,
using the appropriate parity condition.

\noindent
\textbf{Semantic run-trees.} Given a relation $R \, :\,   \colorbang X \otimes \colorbang A \multimap A $ and $a \in A$,
we define the set $\runtree{R}{a}$ of \emph{semantic run-trees} of $R$ producing $a \in A$
as the set of possibly infinite $\left(X \uplus A\right)$-labelled trees,
with nodes colored by elements of $Col$, and such that the four conditions below are satisfied:
\begin{enumerate}
\item the root of the tree is labelled by $a$, and has neutral color $\epsilon$,
\item the inner nodes of the tree are labelled by elements of the set $A$,
\item the leaves are labelled by elements of the set $X \uplus A$,
\item for every node labelled by an element $b \in A$:
\begin{itemize}
\item if $b$ is an inner node, letting $a_1, \cdots, a_n$ denote the labels 
of its children belonging to $A$ and $x_1, \cdots,\, x_m$ the labels belonging to $X$:
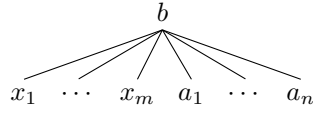
$$
\begin{tikzpicture}
\Tree [.$b$ $x_1$ $\cdots$ $x_m$ $a_1$ $\cdots$ $a_n$ ]
\end{tikzpicture}
$$
and letting $c_i$ (resp. $d_j$) be the color of the node labelled $x_i$ (resp. $a_j$),
$$
(\left\{(c_1,\,x_1),\cdots,\,(c_m,x_m)\right\},\,\left\{\,(d_1,\,a_1),\cdots ,\,(d_n,\,a_n)\right\},b) \in R
$$
\item if $b$ is a leaf, then $(\emptyset,\,\emptyset,\,b) \in R$.
\end{itemize}
\end{enumerate}

\noindent
At this point, we adapt to semantic run-trees the usual acceptance condition 
on the run-trees of an alternating parity automata:
an infinite branch of the semantic run-tree is winning if and only if 
an element of $Col \setminus \{\epsilon\}$ occurs
infinitely often along it, and if the maximal such element is even.
A semantic run-tree is declared winning if and only if all its infinite branches are.\\

\noindent
Given a semantic run-tree $\tree$, we define the set $\leaves{\tree} \subseteq \colorbang X$
as the set of elements $(c,x)$ where $(c',x)$ is a leaf of $\tree$ labelled with $x \in X$, 
and $c$ is the maximal color encountered on the path from the leaf to the root or $\tree$.\\

\noindent
\textbf{Fixpoint operator.} We now define the fixpoint of a binary relation 
$$
R \, :\,  \colorbang X \otimes \colorbang A \multimap A
$$
as the downward-closed binary relation
\begin{equation}\label{master-equation}
\begin{tabular}{rclr}
\hspace{-.8em}$\fixpoint{X,A}\,(R)$ & $\, = \,$ & $\{ \, (u,\,a) \, | \,$ &  $\exists \tree \in \runtree{R}{a} \,\, \mbox{with} \,\, u = \leaves{\tree} $\\
&&& $ \mbox{and } \tree \mbox{ is a winning semantic run-tree.} \, \}$\\
\end{tabular}
\end{equation}

\begin{proposition}
The fixpoint operator $\fixpoint{}$ is a Conway operator over $\FinScott$.
%the full subcategory of $\Scott$ consisting of finite ordered sets.
%
Consequently, its Kleisli category $\FinScottcol$ is a model
of the $\lambda Y$-calculus.
\end{proposition}

\noindent
As in \S\ref{section/type-denotations}, we find useful and even illuminating 
to formulate a type-theoretic counterpart to our definition of the Conway operator $\fixpoint{X,A}$
provided by the following typing rule $Y_{\sigma}$
which should be added to the original type system of Figure~\ref{fig/typesys} :
% -- note that the purpose of the parameter $X$ is to handle contexts:\\

\quad

\noindent\makebox[\textwidth][c]{
$$
\AxiomC{$\Gamma_0 \ \vdash \ M\,:\,\{\,(c_1,\, \beta_1),\,\ldots,\,(c_n,\, \beta_n)\,\} \rightarrow \alpha\,::\,\sigma \rightarrow \sigma$}
\AxiomC{$\Gamma_i \ \vdash \ Y_{\sigma}\,M\,:\,\beta_i\,::\,\sigma$}
\LeftLabel{$Y_{\sigma}$ \ }
\BinaryInfC{$\Gamma_0 \cup \square_{c_1}\, \Gamma_1 \cup \cdots \cup \square_{c_n} \, \Gamma_n \ \vdash \ Y_{\sigma}\ M \,:\,\alpha\,::\,\sigma$}
\DisplayProof
$$
}\ \\

\noindent
In the resulting intersection type system,
derivations of infinite depth are allowed, and have colored nodes,
defined as follows:
\begin{itemize}
\item for every occurrence of the rule $Y_{\sigma}$, we assign color $c_i$ to the node $\Gamma_i \ \vdash \ Y_{\sigma}\,M\,:\,\beta_i\,::\,\sigma$.
\item all the other nodes are assigned the neutral color $\epsilon$.
\end{itemize}
An infinite derivation tree is then accepted as a proof of the system
when all its branches are winning, in the same sense as for the branches
of a semantic run-tree.

\begin{theorem}\label{th/typage-semantique}
Given a $\lambda Y$-term $t$, the sequent 
$$
\Gamma\ =\ x_1\,:\,u_1\,::\,\sigma_1,\,\ldots,\,x_n\,:\,u_n\,::\,\sigma_n \ \vdash \ t\,:\,\alpha\,::\,\tau
$$
has a winning derivation tree in the type system with fixpoints iff
$$
(u_1,\,\ldots,\,u_n,\,\alpha) \ \in \ \interpA{\Gamma\ \vdash\ t\,::\,\tau}  \ \subseteq \ \left(\,\colourbang\,\interpcol{\sigma_1} \otimes \cdots \otimes \colourbang \interpcol{\sigma_n}\,\right) \multimap \interpcol{\tau}
$$
%where the denotation is computed in $\Scottcol$ enriched with the parameterized fixpoint operator $\fixpoint{X,A}$.
\end{theorem}
\noindent
At this point, we take advantage of the correspondence recalled in Proposition~\ref{prop/hors-lambdaY}
between higher-order recursion schemes (HORS) on the ranked alphabet~$\Sigma$,
and closed $\lambda Y$-terms with constants in the same alphabet~$\Sigma$.
Indeed, the correspondence enables us to justify the following typing rule
for HORS :

\quad

\noindent\makebox[\textwidth][c]{
$$
\AxiomC{$\Gamma_0,\,F\,:\,\{\,(c_1,\, \beta_1),\,\ldots,\,(c_n,\, \beta_n)\,\}\,::\,\sigma \ \vdash \ \mathcal{R}(F)\,:\, \alpha\,::\,\sigma$}
\AxiomC{$\Gamma_i \ \vdash \ F\,:\,\beta_i\,::\,\sigma \quad (\forall\,i)$}
\BinaryInfC{$\Gamma_0 \cup \square_{c_1}\, \Gamma_1 \cup \cdots \cup \square_{c_n} \, \Gamma_n \ \vdash \ F \,:\,\alpha\,::\,\sigma$}
\DisplayProof
$$
}\\

\noindent
which provides a direct mean to type the HORS $\mathcal{G}$ 
in the intersection type system, in such a way as to reflect
its interpretation $\interpA{\G}\subseteq Q$ in the Scott semantics.

%%
%For every non-terminal $F \in \mathcal{N}$ of simple type $\sigma$,
%and every context $\Gamma$, we define
%$$
%\interpA{\Gamma\ \vdash\ F}\ \ =\ \ \interpA{\Gamma\ \vdash\ Y_{\sigma}\,\left(\, \lambda F.\,\mathcal{R}(F)\,\right)}
%$$
%%
%%We then define the semantics of a HORS $\G$ by setting $\interpcol{\G}\,=\,\interpcol{S}$. 
%%
%This translation motivates the introduction of the typing rule\\
%
%\noindent\makebox[\textwidth][c]{
%$$
%\AxiomC{$\Gamma_0,\,F\,:\,\{\,(c_1,\, \beta_1),\,\ldots,\,(c_n,\, \beta_n)\,\}\,::\,\sigma \ \vdash \ \mathcal{R}(F)\,:\, \alpha\,::\,\sigma$}
%\AxiomC{$\Gamma_i \ \vdash \ F\,:\,\beta_i\,::\,\sigma \quad (\forall\,i)$}
%\BinaryInfC{$\Gamma_0 \cup \square_{c_1}\, \Gamma_1 \cup \cdots \cup \square_{c_n} \, \Gamma_n \ \vdash \ F \,:\,\alpha\,::\,\sigma$}
%\DisplayProof
%$$
%}\\
%
%\noindent
%which allows to type of the rewrite rules of HORS. We define the semantics of a recursion scheme $\G$ of start symbol $S$ as $\interpA{\G}\,=\,\interpA{\,\,\vdash\, S :: o\,\,}$.
%\begin{corollary}
%\end{corollary}

\section{Finitary semantics solve the selection problem}
\label{section/selec}

The first theorem of the section establishes a perfect correspondence
between our finitary interpretation $\interpA{\G}$ of the higher-order recursion scheme $\G$
in the Scott semantics, and the set of accepting states of the automaton~$\APT$ :

\begin{theorem}\label{th/scott-model-checking}
An alternating parity tree automaton $\mathcal{A}$ has 
an accepting run-tree with initial state $q_0$ over the value tree $\valG$ of a
higher-order recursion scheme $\G$ if and only if $q_0 \in \interpA{\G}$.
\end{theorem}

By Theorem~\ref{th/typage-semantique}, checking whether $q_0 \in \interpA{\G}$ is equivalent
to checking whether there exists a derivation of the sequent $\emptyset \,\vdash\,S\,:\,q_0\,::\,o$ 
in the colored intersection type system
defined in \S\ref{section/rec}.
Since the interpretation of simple types in $\FinScott$ is finite,
% -- and thus the sets of intersection types refining them -- 
only finitely many intersection types and contexts may occur in such a derivation.
Hence, searching for a derivation of the sequent $\emptyset \,\vdash\,S\,:\,q_0\,::\,o$
reduces in this case to solving a finite parity game whose nodes are precisely
the sequents of the derivation tree.
This has the following immediate consequence:

\begin{corollary}
The local model-checking problem is decidable.
\end{corollary}

Recall moreover that the existence of a winning strategy 
in a finite parity game implies that there exists a \emph{memoryless} winning strategy.
In this setting, winning strategies correspond to winning derivation trees of the intersection type system,
and memoryless strategies correspond to derivation trees admitting a finite representation
using backtracking pointers.
From such a finite representation~$\pi$, one can define a higher-order recursion scheme $\G_q$
on a ranked alphabet~$\Sigma_{\APT}$ obtained from $\Sigma$ by annotating every terminal $a$
with elements of its interpretation $\interpA{\,a\,}$.
The HORS $\G_q$ has a non-terminal $F_{\alpha}(o)$ for every occurrence~$o$
of the non-terminal $F$ in the finite representation $\pi$ of the derivation tree,
where $\alpha$ is the intersection type of the occurrence~$o$ of $F$ in $\pi$.
Each occurrence~$o$ of a non-terminal~$F$ then induces a rewrite rule $F_{\alpha}(o)\to_{\G_q} term(o)$
where $term(o)$ is an annotated version of the $\lambda$-term $\mathcal{R}(F)$
coming from the original scheme $\G$.
The annotation of $term(o)$ is obtained by annotating the non-terminals
and the terminals of $\mathcal{R}(F)$ with the intersection types occurring
in the finite representation~$\pi$ of the derivation tree.
%
%Finally, the backtracking pointers are replaced in $\pi$ with calls to a previously introduced non-terminal.
%
This defines a higher-order recursion scheme $\G_q$, which generates a run-tree 
$\valuetree{\G_q}$ of the alternating parity tree automaton $\APT$
over $\valG$.
%, with the mild difference that terminals of $\Sigma$ are annotated with transitions, and not only with states.
As a consequence:

\begin{theorem}
The selection problem is decidable.
\end{theorem}
%
%\vspace{-2em}
%

%\section{Related works}
%\label{section/related-works}
%
\section{Conclusions and perspectives}
\label{section/ccl}
In this paper, we explain how to apply our semantic approach to higher-order model-checking
based on linear logic, in order to establish the decidability of local model-checking
and of the selection problem.
Our approach provides a rigorous and compositional approach to higher-order model-checking,
and adapts to the inductive-coinductive framework of MSO logic a nice and well-established
connection between linear logic,  Scott domains, and intersection types.
Future work includes a detailed comparison with a similar line of work on finite models
of the $\lambda\,Y$-calculus currently developed by Salvati and Walukiewicz~\cite{salvati-walu-finite-model}.
\bibliography{model-theoretic-selection}
\bibliographystyle{splncs03}
%\end{footnotesize}

%\newpage

%\appendix
%\section*{App}

\end{document}